\documentclass[aps,prb,twocolumn,showpacs]{revtex4-1}%
\usepackage{amsfonts}
\usepackage{amsmath}
\usepackage{amssymb}
\usepackage{graphicx}
\usepackage{dcolumn}
\usepackage{bm}

\begin{document}

\title
{ Elastic fields in superconductors  caused by moving vortices
}
\author{V. G. Kogan}
\email{kogan@ameslab.gov}

\affiliation{  Ames Laboratory, Ames, Iowa 50011, USA  }
 \date{\today} 

 \begin{abstract}
 Strains in superconductors due to 
 moving vortices and vortex lattices are discussed. It is shown that the energy stored in  elastic strains increases with vortex velocity. For moving vortex lattices, the elastic energy depends on velocity orientation relative to the lattice having minimum for the velocity directed along one of the unit cell vectors of the moving  lattice. It is shown that for supersonic motion, the vortex induced stress field has a shape similar to  supersonic shock waves.
 \end{abstract}
\maketitle
 
\section{Introduction}

The subject of this work are vortex induced strains in 
superconductors  which are related to the stress dependence of the critical temperature $\partial T_c /\partial p$.  It turned out   that  this derivative in pnictides, and in Ca(Fe$_{1-x}$Co$_x$)$_2$As$_2$ in particular, \cite{dTc/dp} by one or two orders of magnitude exceeds   values for conventional superconductors, making Fe-based pnictides   favorable for observation of   magneto-elastic effects. 
 
Still, experimental evidence for  strains in the mixed state in the absence of pinning is scarce. In Ref.\,\onlinecite{KBMD}, the experimental disagreement with predicted by the London theory vortex lattice structure in NbSe$_2$ in tilted fields was attributed to extra magneto-elastic interactions of vortices. Such comparisons are difficult because complete sets of  elastic moduli are usually unavailable.  Besides, vortex lattices are extremely sensitive to multitude of factors such as electronic band structure, order parameter symmetry, etc. 

The stress  due to vortices should affect the free energy of the system proportional to their number, i.e. to the magnetic induction $B$, and--along with energy--the equilibrium magnetization $M$.\cite{K2013} This was suggested as a reason for the hump in {\it reversible} $M(B)$ observed in La$_{1.45}$Nd$_{0.40}$Sr$_{0.15}$CuO$_4$, and
CeCoIn$_5$.\cite{Doug,Carmen} The data on reversible $M(B)$ are extremely rare since they imply absence of pinning.

Hence, as it stands today, existence of vortex induced strains and the magneto-elastic intervortex interactions are still to be confirmed. It is hard to expect progress in microscopic description of vortex induced strains at arbitrary temperatures, although near $T_c$ a progress has been made.\cite{Cano1,Cano2}
 For this reason, the London approach within which the core is represented by a delta-function, deserves a try, though clearly it ``sweeps under the rug" many important questions. Nevertheless, given a long history of London approach in describing   magnetic properties of type-II superconductors,  one can hope that applying it to magneto-elastic effects might be useful. 

Magneto-elastic interactions are long-range and, their weakness notwithstanding, should  affect the mixed state in general,\cite{KBMD,Cano1,K2013} intervortex interactions,\cite{K2013a} and   vortex lattice  structures in particular. \cite{Shizeng}  
 In this paper, the focus is on magneto-elastic effects caused by {\it moving} vortices. 
 
 The vortex core of a size $\xi$, the coherence length, is not the only source of elastic strains. Super-currents round the core which extend to distances $\sim\lambda$ are argued to contribute even more to strains becauase usually $\lambda\gg\xi$.\cite{Cano1,Cano2} Since the currents decay exponentially,  the corresponding strain source can be considered as {\it local}, its relatively large size notwithstanding. Hence, at distances $r\gg\lambda$ this source can be considered as point-like, and the approach developed below for  core-size sources can be generalized to $\lambda$-size sources   by a proper rescaling. 

\section{Elastic energy of  vortex at rest}
 
Nucleation of the  normal vortex core strains the host superconductor, since the normal   phase 
has a larger specific volume  as compared to  superconductor. The relative volume change $\zeta$ is related to the pressure dependence of the condensation energy or of the critical field $H_c$: \cite{LLelectro} 
\begin{equation}
\zeta = \frac{V_n-V_s}{V_s}=\frac{H_c}{4\pi}\frac{\partial H_c}{\partial 
p}\,.
\label{zeta}
\end{equation}

 The elastic  energy density in isotropic solids reads: 
\begin{equation}
F=\lambda u_{ll}^2/2 +\mu u_{ij}^2\,. 
\label{1}
\end{equation}
Here, $u_{ij}$ is the strain tensor  and $\lambda$, $\mu$ are 
${\rm Lam\acute{e}}$ coefficients; summation over double indices is implied.\cite{LL} 
The stress tensor $\sigma_{ij} = \partial F/\partial 
u_{ij} = \lambda u_{ll}\delta_{ij} + 2\mu u_{ij}$, and the 
equilibrium condition $\partial \sigma_{ij}/\partial x_j \equiv \sigma_{ij,j} =0$ is    
\begin{equation}
\lambda u_{ll,i} +2\mu u_{ij,j}=0\,. 
\label{3}
\end{equation}
  
For a   single vortex  along $z$ at the origin, the displacement ${\bf u}=(u_x,u_y,0)$ is radial in the plane $xy$, 
i.e., ${\rm curl}\,{\bf u}=0$ or ${\bf u}=\nabla \chi$, and 
$u_{\alpha \beta}=\chi_{,\alpha \beta}$ where $\chi$ is a scalar and $\alpha,\beta$ acquire only $x$ and $y$ values. The equilibrium condition (\ref{3}) reads $(\lambda 
  +2\mu) \chi_{,\alpha \beta \beta }=0 $ with the first integral 
\begin{equation}
\chi_{,\beta \beta } \equiv \nabla^2 \chi = C= {\rm constant}\,.
\label{chi}
\end{equation}
 To fix this constant, one notes that $ \chi_{,\beta \beta }=u_{\beta \beta}$ describes compression and  is related to the hydrostatic pressure within the system. For the problem of the strain caused by a single vortex in otherwise unrestrained crystal, the pressure is zero, and one has to solve $\nabla^2 \chi =0$ under the boundary condition   $u\to 0$ at large distances. Hence, the problem is the same as that of a linear charge in electrostatics, where the potential satisfies 
\begin{equation}
 \nabla^2 \chi = A\delta({\bm r})\,,
\label{electrostat}
\end{equation}
where $A$ is related to the linear charge density. 
From a simple core model as a normal cylinder of a size $\xi$ follows: \cite{KBMD}  
   \begin{equation}
 A=2\pi\xi^2\gamma\,,\quad \gamma = \frac{\zeta(\lambda+\mu)}{2(\lambda+2\mu)}\,. 
 \label{gamma}
\end{equation}
     One, then, has $\chi=( A/2\pi)\ln r +const$ and  
\begin{equation}
{\bm u} = \frac{\gamma  \xi^2{\bm r}}{r^2}\,,\,\,\, 
u_{\alpha \beta} =\frac{\gamma\xi^2}{r^2}\left (\delta_{\alpha 
\beta}-\frac{2}{r^2}x_{\alpha}x_{\beta}\right )\,.
\label{5}
\end{equation}
   
The elastic energy per unit length of a vortex  is:  
\begin{equation}
{\cal E}= \int  d^2\bm r (\lambda u_{\alpha\alpha}^2/2 +\mu u_{\alpha\beta}^2)\,. 
\label{E1}
\end{equation}
According to Eq.\,(\ref{5})   $u_{\alpha\alpha}=0$ and
\begin{equation}
  u_{\alpha\beta}^2=u_{xx}^2+ u_{yy}^2+2u_{xy}^2=\frac{A^2}{2\pi^2r^4} \,. 
\label{uik}
\end{equation}
Integrating over $r$ in Eq.\,(\ref{E1}) from $\xi$ to $\infty$, one obtains:
 \begin{equation}
{\cal E}=\frac{ A^2\mu}{2\pi \xi^2}= 2\pi  \mu \xi^2\gamma^2\,.  
\label{Estat}
\end{equation}

 \section{Vortex moving with subsound velocity}
  
If the vortex moves, the elastic displacement at a particular material point depends on time and the local equation of motion   is $\sigma_{\alpha\beta,\beta}=\rho \ddot u_\alpha$ where $ \ddot {\bm u } \equiv \partial^2{\bm u}/\partial t^2$ and $\rho$ is the material density. 

For a point source of stress  
  one has to add to the energy density a term $\eta_{\alpha \beta}u_{\alpha \beta}\delta(\bm r-\bm v t)$,  so that the stress tensor has an addition  $\partial F/\partial u_{\alpha\beta}=\eta_{\alpha\beta}\delta(\bm r)$.
In isotropic case $\eta_{\alpha \beta}=\eta\,\delta_{\alpha \beta}$ 
and the equation of motion is $(\lambda +
2\mu)\chi_{,\alpha\beta\beta}-\rho\ddot{\chi}_{,\alpha}
=\eta\,\partial_{\alpha}\delta({\bm r}-{\bm v}t)$, with the first integral 
\begin{equation}
(\lambda + 2\mu)\nabla^2\chi -\rho\ddot{\chi} =\eta\,\delta({\bm r}-{\bm v}t)
\label{wave-eq}
\end{equation}
where the coefficient $\eta$ is fixed by comparison with the static Eq.\,(\ref{electrostat}): $ \eta=A(\lambda + 2\mu)$. Hence, we have 
 \begin{equation}
 \nabla^2\chi -\frac{1}{v_s^2}\ddot{\chi} = A\,\delta({\bm r}-{\bm v}t)
\label{wave-eq}
\end{equation}
where $v_s =\sqrt{(\lambda+2\mu)/\rho}$ is the longitudinal sound velocity.\cite{Hydrodynamics}  
 
Equation (\ref{wave-eq}) is  solved by Fourier transform
\begin{equation}
\chi({\bm r},t)=\int \frac{d^2{\bm k}\,d\omega}{(2\pi)^3}\chi({\bm
k},\omega)e^{i({\bm k}{\bm r}-\omega t)}\,.
\label{FT}
\end{equation}
Transforming    the RHS of Eq.\,(\ref{wave-eq}), 
\begin{eqnarray}
 \int_{-\infty}^\infty &dt& \int  d{\bf r}\, \delta({\bm r-\bm v t}) e^{-i({\bm k}{\bf r}-\omega t)}\nonumber\\
 =  \int_{-\infty}^\infty &dt&  e^{-i({\bm k}{\bm v}-\omega) t}
 = 2\pi \delta({\bm k}{\bm v}-\omega)\,,\qquad
\label{delFT}
\end{eqnarray}
one obtains:
\begin{equation}
\chi({\bm k},\omega)= 2\pi  v_s^2A\frac{\delta(\omega-\bm{kv})}{\omega^2 - v_s^2k^2}\,.
\label{khi}
\end{equation}
  In  real space
\begin{equation}
\chi({\bf r},t)=  v_s^2A\int\frac{d^2{\bm k}}{4\pi^2}\,\frac{e^{i{\bm k}({\bf r}-{\bm v}t)}}
{({\bm k}{\bm v})^2-k^2v_s^2 }\,.
\label{chi(r,t)}
\end{equation}
If $v=0$, the static solution is recovered (the integral for $v=0$ is divergent, but the derivative $\partial_r \chi=A/2\pi r$ is the same as in the static case). 

For the velocity $\bm v = v{\hat x}$, one has at $t=0$:   
\begin{equation}
\chi({\bf r},0)=-\frac{A}{4\pi^2 }\int\frac{d^2{\bm k}\,e^{i{\bm k}{\bm r}}}
{(1-V^2)k_x^2+k_y^2}\,,
\label{2}
\end{equation}
where the reduced velocity $V=v/v_s<1$. This differs from the static solution 
  by rescaling $k_x\to \sqrt{1-V^2}\,k_x$ or, in real space, by  $x\to x/\sqrt{1-V^2}$. 
In other words, for $V <1$, the circles of constant $\chi$  for the vortex at rest, become ellipses $x^2/(1-V^2)+y^2=const$ with the $x$ semi-axis   $\sqrt{1-V^2}$ times shorter than that of $y$. Hence,   the static potential $\chi(x,y)$ is 
 contracted  in the direction of motion by the factor
$\sqrt{1-v^2/v_s^2}$ and  moves as a whole with the vortex velocity $v$. 
 
The energy stored in the elastic distortion of a vortex moving with a constant velocity 
is time independent. To evaluate this energy one can use the potential (\ref{2}) for $t=0$. This evaluation can be done in real space. Since the static potential $\chi=(A/4\pi )\ln(x^2+y^2)$, we have for the moving vortex
 \begin{equation}
 \chi =  \frac{A}{4\pi}\,\ln\left(\frac{x^2}{1-V^2}+y^2\right)\,  
\label{lattice}
\end{equation}
(an irrelevant constant is omitted).  
Clearly, the lines of constant $\chi$ (the circles for the vortex at rest) are elliptic with a short semi-axis $ \sqrt{1-V^2}$ along the direction of motion; this ellipse is strongly squeezed when the velocity approaches that of the sound. The strains follow:
\begin{eqnarray}
 u_{xx}&=&-  \frac{A(x^2-\beta^2y^2)}{2\pi(x^2+\beta^2y^2)^2}  \, ,\quad  u_{yy}=   \frac{A\beta^2(x^2-\beta^2y^2)}{ 2\pi(x^2+\beta^2y^2)^2} , \nonumber\\
 u_{xy}&=&-  \frac{ A\beta^2 xy}{ \pi(x^2+\beta^2y^2)^2}   \,,\qquad \beta^2=1-V^2 \,.
\label{strains}
\end{eqnarray}

  The elastic energy (\ref{E1}) can now be evaluated as shown in Appendix A:
 \begin{eqnarray}
{\cal E}=   \frac{  A^2 (1+\beta^2)}{16\pi\xi^2\beta^3}\left[\frac{\lambda}{2}(1- \beta^2 )^2+\mu(1+\beta^2)^2\right] .
  \label{E_mov_vort}
\end{eqnarray}
 
This gives for low velocities $V=v/v_s\ll1$:
 \begin{eqnarray}
 {\cal E}\approx  \frac{ A^2 \mu}{ 2\pi\xi^2 }\left(1+\frac{  \lambda+3\mu  }{8\mu}V^4\right)  
  \label{smallV}
\end{eqnarray}
with correct limit (\ref{Estat}) for $V=0$. 
If the velocity approaches $v_s$, the elastic energy diverges as
\begin{eqnarray}
{\cal E}\approx  \frac{A^2  }{16 \sqrt{2}\pi \xi^2} \frac{\lambda+2\mu}{(1-V)^{3/2}}  \,.
  \label{v->vs}
\end{eqnarray}
 
The elastic potential $\chi(\bm r,t)$ obtained  solving  linear Eq.\,(\ref{wave-eq}) is a partial 
solution of this equation with the source term $\propto \delta(\bm r -\bm v t)$ at the RHS. In fact, this equation has also solutions of the homogeneous  equation without the RHS, i.e. of the wave equation $\nabla^2\chi -\ddot{\chi}/v_s^2 = 0$. These are sound waves generated by the moving elastic field; these waves carry away  energy and contribute to the vortex drug coefficient. This problem, however, is out of the scope of this paper. 
 
 \section{Vortex lattice}
 \subsection{Static lattice}
 
Consider now a 2D periodic lattice of vortices at positions $\bm a$ in an infinite sample.     
As argued in Ref.\,\onlinecite{K2013}, the elastic potential in this case is a solution of
 \begin{equation}
\nabla^2\chi = A\left[\sum_{\bm a}\delta(\bm {r-a})-\frac{B}{\phi_0}\right]\,, 
\label{lattice1}
\end{equation}
where $B$ is the magnetic induction.    In terms of electrostatic analogy, the source term of the Poisson equation (\ref{lattice1}) must have a  negative background density $B/\phi_0$  to make the system ``quasi-neutral" for the equation to have periodic finite solutions on the whole plane.

 One looks for $\chi(\bm r)$ as   Fourier series
 \begin{equation}
 \chi (\bm r)=  \sum_{\bm G}\chi(\bm G)e^{i{\bm G}{\bm r}},\,\,\,\,\chi(\bm G)=
 \frac{B}{\phi_0} \int d{\bm r}\chi(\bm r)e^{-i{\bm G}{\bm r}}\qquad 
\label{11}
\end{equation}
with $\bm G$ being the reciprocal lattice vectors. 
Transforming the RHS of   Eq.\,(\ref{lattice1}) one can use   identities:\cite{LLstat} 
  \begin{eqnarray}
   \sum_{\bm a} \delta(\bm r-\bm a)=\frac{B}{\phi_0}\sum_{\bm G}
   e^{i{\bm G}{\bm r} },\quad \frac{B}{\phi_0}=\frac{  B}{\phi_0} \sum_{\bm G} \delta_{{\bm G},0} 
 e^{ i {\bm G}{\bm r} } ,\qquad
   \label{identity}
\end{eqnarray}
where the symbol $\delta_{0,0}=1$ and $\delta_{\bm G,0}=0$ for $\bm G\ne 0$. 
One then obtains:
\begin{equation}
\chi(\bm G) = - \frac{2\pi \gamma\xi^2B}{\phi_0G^2} \left(1-\delta_{\bm G,0}\right)\, .
\label{chi(G)}
\end{equation}
Being transformed to real space, the part of $\chi(\bm G)$ containing $\delta_{\bm G,0}$ generates an uncertain constant, 
\begin{equation}
\sum_{\bm G}   \frac{1-\delta_{\bm G,0}}{ G^2}\,e^{i{\bm G}{\bm r}}  = \sum_{\bm G\ne 0}   \frac{e^{i{\bm G}{\bm r}}}{ G^2}+ const\, ,
\label{chi(r)'}
\end{equation}
which is   coordinate independent  and therefore irrelevant. 

\subsection{Moving lattice}

If the lattice moves, the vortex positions are $\bm a+\bm v t$ where $\bm a$ define the lattice at $t=0$. The   elastic potential obeys 
 \begin{equation}
\nabla^2\chi -\frac{\ddot{\chi}}{v_s^2}= A\left[\sum_{\bm a}\delta({\bm r}-{\bm v}t-{\bm a})-\frac{B}{\phi_0}\right]\,.
\label{lat-mov}
\end{equation}

One now employs the Fourier transform 
 \begin{eqnarray}
 \chi (\bm r,t)&=&  \sum_{\bm G}\int \frac{d\omega}{2\pi}\chi(\bm G,\omega)e^{i({\bm G}{\bm r}-\omega t)},\label{FTb}\\
  \chi(\bm G,\omega)&=&
 \frac{B}{\phi_0} \int d{\bm r}\,dt\,\chi(\bm r,t)e^{-i({\bm G}{\bm r}-\omega t)} \,,  
\label{FTc}
\end{eqnarray}
see   Appendix B, to obtain:
 \begin{eqnarray}
  \chi(\bm G,\omega)&=&
 \chi_0 \,\frac{\delta(\omega-{\bm G}{\bm v}) -\delta(\omega)\delta_{\bm G,0}}{\omega^2-G^2v_s^2} \label{chi-v}\\    
  \chi_0&=& \frac{4\pi^2B\gamma\xi^2v_s^2}{\phi_0}=\frac{2\pi B \gamma v_s^2}{H_{c2}}\,.
 \label{chi0}
\end{eqnarray}
Hence, only the frequencies $\omega={\bm G}{\bm v}$ are present in the  Fourier transform over time.\cite{Lev} One could expect this because an observer of a moving vortex lattice should register the main period $a_1/v\propto 1/G_1v$ where $a_1$ is the real space lattice period and $G_1$ is the corresponding reciprocal vector.

 \subsection{Elastic energy of moving lattice}
 
The quasi-static elastic energy density of the vortex lattice is:
\begin{equation}
F=\frac{B}{\phi_0}\int_{\rm cell} d\bm r (\lambda u_{\alpha\alpha}^2/2 +\mu u_{\alpha\beta}^2)\,, 
\label{1}
\end{equation}
where   the integration is extended over the unit cell. 
 
Being interested in a stationary state, one can calculate the energy at $t=0$.    Fourier component of the potential $\chi$ at $t=0$ are 
 \begin{eqnarray}
  \chi(\bm G, 0)&=& \int_{-\infty}^\infty \frac{d\omega}{2\pi}  \chi(\bm G,\omega)\nonumber\\
  & =& \frac{\chi_0}{2\pi} \left[\frac{1}{ ({\bm G}{\bm v})^2 - G^2v_s^2 } +\frac{\delta_{\bm G,0}}{G^2v_s^2} \right] \,. \qquad 
\label{chi(G,0)}
\end{eqnarray}
 As in the static case, the last term here generates in real space a coordinate independent constant which can be disregarded because only gradients of $\chi$ have physical meaning. 

After straightforward algebra one obtains:
 \begin{eqnarray}
  F&=&\frac{\chi_0^2(\lambda+2\mu)}{2}  \sum_{\bm G\ne 0}
   \frac{G^4}{ [({\bm G}{\bm v})^2 - G^2v_s^2]^2 }   \,.  
\label{F}
\end{eqnarray}
The sum here is divergent  due to  long-range elastic perturbations. It   can be evaluated numerically because, when  treating vortex cores as $\delta$-functions, the maximum $G$ should be of the order of  $1/\xi$. 
 
One can now compare energies of hexagonal vortex lattices moving with velocities $\bm v$ oriented differently relative to the lattice. 
Let the hexagonal lattice  have one of the unit cell vectors along the $x$ axis:
 \begin{eqnarray}
  \bm a_{1}=a_0 {\hat {\bm x}} ,\,\,  \bm a_{2}=a_0( {\hat {\bm x}}+\sqrt{3}{\hat {\bm y}})/2,\,\, a_0^2=2\phi_0/B\sqrt{3}  , \qquad 
\label{lat-real }
\end{eqnarray}
that correspond to the reciprocal lattice
 \begin{eqnarray}
 G_x=G_0\frac{ \sqrt{3}}{2}\,n ,\,\,  G_y=G_0\left(  m-\frac{n}{2}\right),\,\,  G_0^2=\frac{2B}{ \phi_0\sqrt{3}}  \qquad 
\label{lat-G}
\end{eqnarray}
with integers $n$ and $m$. The squared length of a   lattice vector is $G^2=G_0^2(n^2-nm+m^2)$. 
The velocity orientation is fixed by the angle $\alpha$ with the $x$ axis: $\bm v = v ({\hat x}\cos\alpha +{\hat y}\sin\alpha)$. Hence, we have for the sum in Eq.\,(\ref{F}): 
  \begin{eqnarray}
   S(\bm v)&=& v_s^4 \sum_{\bm G\ne 0}
   \frac{G^4}{ [({\bm G}{\bm v})^2 - G^2v_s^2]^2 } \nonumber\\
  &=&  \sum_{n^2+m^2\ne 0} \frac{(n^2-nm+m^2)^2 }{d(\alpha,n,m)}\,,\label{S} \\
    d&=&\Big[V^2 \left(  \sqrt{3}n \cos\alpha+(2 m- n )\sin\alpha\right)^2/4\nonumber\\
&-&(n^2-nm+m^2)\Big]^2    
\nonumber
\end{eqnarray}
where $V=v/v_s$.  
To ensure that the sum is not extended to $G \gtrsim 1/\xi$, we introduce in the numerator of the sum (\ref{S}) a factor 
  \begin{eqnarray}
\exp\left(- G^2\xi^2\right)& =&\exp\left[- h  (n^2-nm+m^2)\right],\qquad \nonumber\\
h &=&\frac{2B\xi^2}{\phi_0\sqrt{3}} \sim \frac{ B }{H_{c2}} .
\label{exp} 
\end{eqnarray}

\begin{figure}[h ]
\begin{center}
 \includegraphics[width=8.cm] {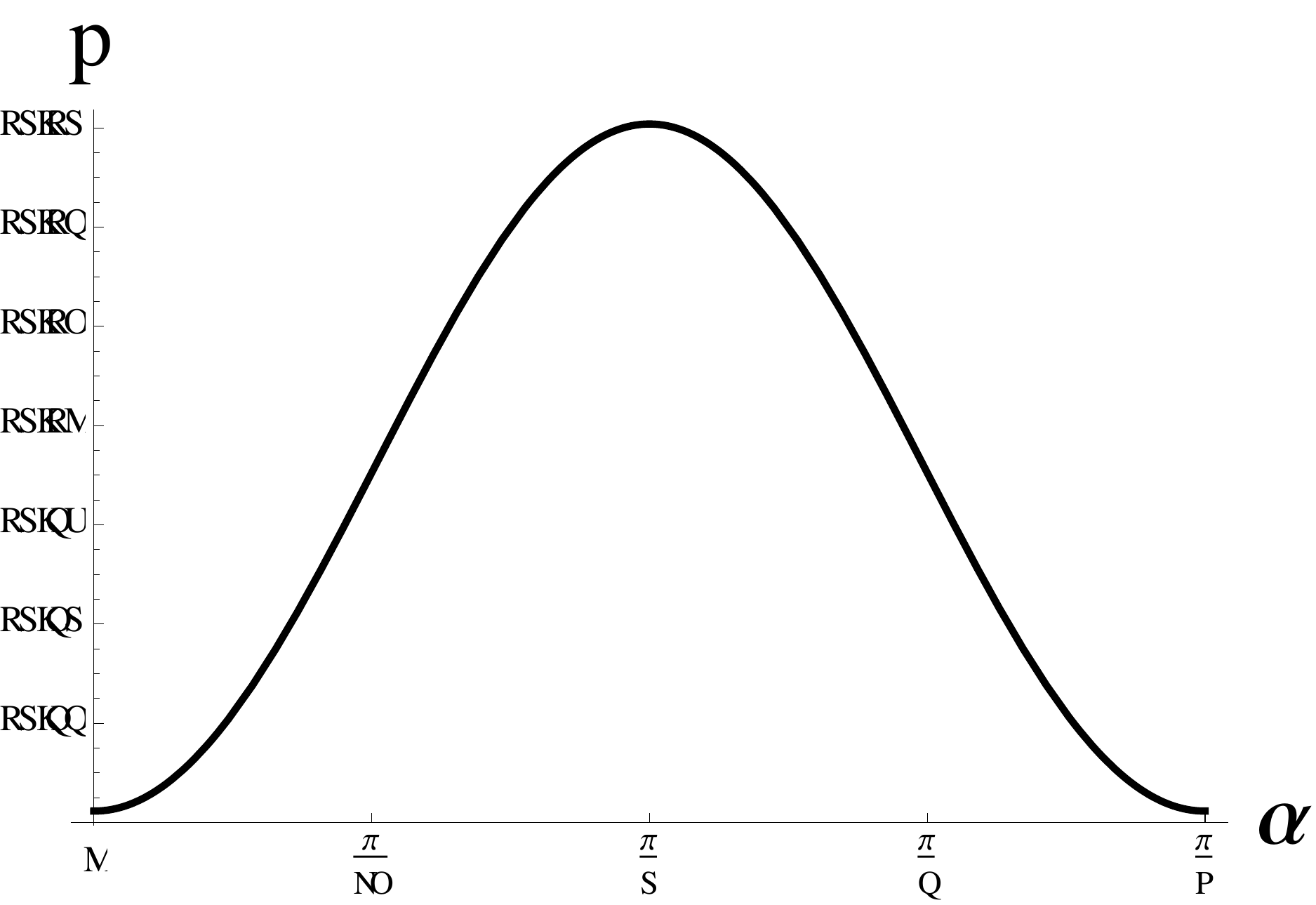}
\caption{  $S(\alpha)$ of Eq.\,(\ref{S}) calculated for  $h=0.1$ and $V=v/v_s=0.6$.  }
\label{f1}
\end{center}
\end{figure}

Figure \ref{f1} shows the  numerically evaluated $S(\alpha)$ according to Eqs.\,(\ref{S}) and (\ref{exp}) for   $v=0.6 \,v_s$ and $h=0.1$ (or $B/H_{c2}\approx 0.5$). It is seen that minimum energy corresponds to velocity directed along the unit cell vectors, i.e. to $\alpha=0$ and $\alpha=\pi/3$, whereas $\alpha=\pi/6$ corresponds to the maximum energy. Although the difference $S(\pi/6)-S(0)$ is of the order $10^{-3}$, it is worth recalling that in the isotropic case the hexagonal vortex lattice at rest is degenerate relative to arbitrary rotations. The difference $\Delta S=S(\pi/6)-S(0)$ increases fast with increasing velocity. 

The difference in elastic energies for the two orientations is
  \begin{eqnarray}
\Delta F \approx  \frac{\chi_0^2\tilde\lambda}{v_s^4}  \Delta S \approx \left(\frac{B}{H_{c2}}\gamma\right)^2 \tilde\lambda \Delta S\,,   
\end{eqnarray}
where $ \tilde\lambda\sim 10^{12}\,$egr/cm$^3$ is an estimate for the elastic constants. Taking $\gamma\sim \zeta \sim10^{-5}$, one obtains  $\Delta F\sim 10^{-3}$erg/cm$^3$ for our example. 

\section{Supersound velocities}

    Velocities of vortices up to $ \approx 1.5\times 10^6\,$cm/s have been recently recorded in Pb films,\cite{Zeldov} which is well above the sound speed in Pb of about  $2\times 10^5\,$cm/s. Vortices were crossing the narrow part of a thin-film bridge and were visualized with the help of submicron-size scanning SQUID. The voltage and the current along the bridge were monitored that made it possible to evaluate vortex velocities. Vortices enter the bridge as a well formed and stable chain but slow down penetrating the bridge (the driving current   decreases toward the strip middle) and at some points the chain splits  in two  or more parallel chains. 
      One may speculate that the  elastic perturbations caused by fast moving vortices   play a role in forming  vortex chains. 

If a vortex moves with a supersound velocity $V>1$, Eq.\,(\ref{2}) for the elastic potential at $t=0$ becomes
\begin{equation}
-\frac{4\pi^2\chi({\bf r},0)}{A}= \int\frac{d^2{\bm k}\,e^{i{\bm k}{\bm r}}}
{k_y^2-\eta^2 k_x^2 },\,\,\, \eta^2=V^2-1>0.  
\label{e40}
\end{equation}
 The integration over $k_y$ is done in the complex plane of $k_y$. The contour of integration for $y>0$ is chosen as a half-circle of a large radius in the upper half-plane, the real axis of $k_y$, and infinitesimal half-circles round the poles at $k_y=\pm \eta k_x$ chosen as to leave the poles within the contour:  
 \begin{equation}
  \int_{-\infty}^\infty\frac{dk_y\,e^{i k_y y}}
{k_y^2-\eta^2 k_x^2 }=-\frac{y}{|y|}\frac{\pi\sin(k_x\eta y)}{\eta k_x}\,.
\label{e41}
\end{equation}

\begin{figure}[h ]
\begin{center}
 \includegraphics[width=8.cm] {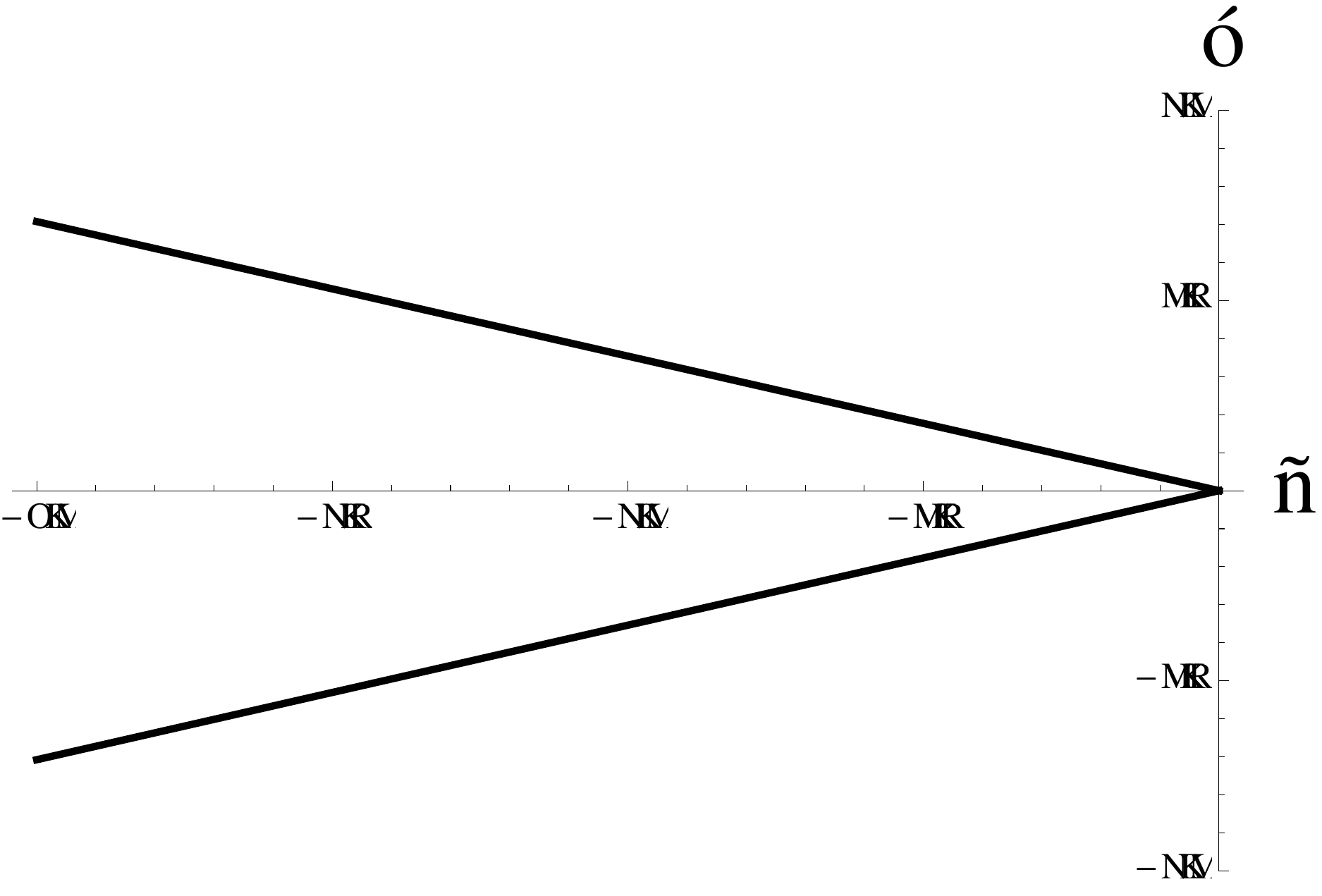}
\caption{Boundaries at which the potential $\chi$ changes discontinuously for $V=v/v_s=3$. $\chi=0$ between these boundaries and $\chi=A/4\eta$ otherwise.
}
\label{f2}
\end{center}
\end{figure}

Integration over $k_x$ then gives:
\begin{equation}
\chi({\bf r},\eta)_{t=0}     = \frac{A}{4 \eta }\, \theta(x+\eta |y|) \,,
\label{e42}
\end{equation}
the  $\theta$-function here is unity for a positive argument and zero otherwise. 
The potential $\chi(x,y)$ should be determined   for $x<0$, because no elastic perturbation can exist in front of a vortex moving with supersound velocity. In fact,  the potential $\chi(x,y)=A/4\eta$ or zero,   depending on what domain of $x,y$ is chosen. These domains are separated by straight lines $ x+\eta y =0$ and $x-\eta y =0$, as shown in Fig.\,\ref{f2}.
The angle between  two boundaries is
\begin{equation}
\alpha   =2\cot^{-1} \eta   \,.
\label{e43}
\end{equation}
In experiment of Ref.\,\onlinecite{Zeldov} $v/v_s\approx 15/2=7.5$ that corresponds to the opening angle of the supersonic tale  behind the vortex  $\alpha\approx 0.27\approx 15^\circ$.

When $v\to v_s$ and $\eta\to 0$, $\alpha\to \pi$, i.e. boundaries of the central domain open to coincide with the $y$ axis. At large velocities $\eta\gg 1$ and $\alpha\approx 2/ \eta \ll  1$. This two-dimensional picture is reminiscent of shock  wave fronts created by a supersound motion of a body in continuous medium. 

Within London model,  the potential $\chi$ changes discontinuously at these boundaries, i.e., the displacement $\bm u$ has a $\delta$-function singularity at the boundaries.     These discontinuities are the model artifacts  since the vortex   is considered as a point source  of the elastic perturbation and is represented by  $\delta$-function. Clearly, in a better theory the discontinuity should be smeared over a belt of a width $\sim\xi$ (or $\lambda$), and the displacement will have a sharp maximum at the boundary and go to zero out of the belt. In turn, the strains related to second derivatives of $\chi$ will have a sharp maximum on one side of the boundary and a sharp minimum on the other. Hence, as a consequence of a sharp displacement at the boundary, one expects the pressure enhancement on one side and the depression on the other. Interaction of other vortices with these pressure profiles is an open question.

 \section{Summary and Discussion}

It is shown that the rotational degeneracy of the hexagonal vortex lattice is removed by lattice motion due to vortex induced strains.   Moreover, the orientation of the moving lattice with one of the  unit cell vectors along the velocity corresponds to  minimum elastic energy. It is worth noting here that the same orientational effect has been predicted by considering pinning, \cite{Kosh-Vin}  the time-dependent Ginzburg-Landau or London models of moving vortex lattices. \cite{Barukh,TDL} 
 
In addition to reorientation of the moving lattice, elastic contribution to the intervortex interaction should cause distortions of the moving lattice structure as compared to   the lattice at rest. Distortions of static lattices in  tetragonal crystals were discussed in Refs.\,\onlinecite{Shizeng,K2013}. The stationary structure of a uniformly moving lattice can be found employing the minimum dissipation principle,\cite{TDL}  the problem which would take us out of the main subject of this paper.
 
  Experimentally, there are situations when the vortex
velocities are very high.  In flux avalanches   in thin YBCO films, vortices are claimed to move with velocities up to $5 \times 10^6$\,cm/s. \cite{Lederer}  In Pb films
recently recorded vortex velocities reach $10^6\,$cm/s.
  \cite{Zeldov}  Hence,  vortex velocities may approach and exceed  $v_s$,  the speed of sound. 
Within the model of this paper, however, the elastic energy diverges when  $v \to v_s$. Clearly, the model   should break down in this limit, because the energy of the vortex system cannot exceed the superconducting condensation energy. 
In other words, the superconducting phase should be strongly perturbed in this case, that would  lead to velocity dependence of basic material parameters  such as $\zeta, \xi, A$, considered as constants in this paper.  

As mentioned, the elastic potential $\chi(\bm r)$ obtained  solving  Eq.\,(\ref{wave-eq}) is only  a partial solution corresponding to  $\delta(\bm r -\bm v t)$ at the RHS. In fact, this equation has also solutions  for zero RHS, i.e. of the wave equation $\nabla^2\chi -\ddot{\chi}/v_s^2 = 0$. These are sound waves which contribute to energy dissipation, i.e., to vortex drag coefficient. Of a special interest the sound generation is for a supersound velocities,
  the subject of future work. 

The density difference between the normal core and the superconducting surrounding is not the only reason for vortex-induced elastic perturbations discussed above. The non-uniform distribution of supercurrents round the vortex may also cause elastic distortions in the underlying crystal. The source  
of these distortions is localized in a region of a size $\lambda$, the London penetration depth which is usually large relative to the core size $\xi$.\cite{Cano1,Cano2} Elastic perturbations, though, decay only as a power law at $r> \lambda$, i.e., they are long range, hence their localized source can again be  formally described by a $\delta$-function. Hence, in principle, the formal treatment presented above can be applied in this case, too. The factor $A$ of the    $\delta$-function will, of course, differ from that given in Eq.\,(\ref{electrostat}). \\
 
I thank  A. Gurevich, E. Zeldov, R. Mints, and L. Boulaevskii for discussions and critical comments.
This work was supported by the U.S. Department of Energy (DOE), Office of Science, Basic Energy Sciences, Materials Science and Engineering Division. The work was done at the Ames Laboratory, which is operated for the U.S. DOE by Iowa State University under contract \# DE-AC02-07CH11358. 
 
 \appendix
 
 \section{ }

To evaluate the energy (\ref{E1})    one needs:
\begin{eqnarray}
 &&(u_{\alpha\alpha})^2 =
  \frac{A^2 (\beta^2-1)^2(x^2-\beta^2y^2)^2}{4\pi^2(x^2+\beta^2y^2)^4} \, \,,\label{u_aa^2}\\
&&u_{\alpha\gamma}^2 = u_{xx}^2+u_{yy}^2+2u_{xy}^2 \nonumber\\
 && =  \frac{A^2 (1+\beta^4)   (x^4+\beta^4y^4) - 2x^2y^2\beta^2(1-4\beta^2+\beta^4)}{4\pi^2(x^2+\beta^2y^2)^4}. \qquad
\label{u_aa, uab^2}
\end{eqnarray}
Further, one has:
\begin{eqnarray}
&&I_1=  \int d{\bm r} u_{\alpha\alpha}^2 =\nonumber\\
&& \frac{A^2 (1-\beta^2) ^2}{4\pi^2} \int_\xi^\infty\frac{dr}{r^3}\int_0^{2\pi} d\varphi \frac{(\beta^2\sin^2\varphi-\cos^2\varphi)^2 }{(\cos^2\varphi+\beta^2\sin^2\varphi)^4}  \qquad\nonumber\\
 &&= \frac{  A^2(1-\beta^2)^2(1+\beta^2)}{16\pi\xi^2\beta^3} \,.
  \label{int1}
\end{eqnarray}
Further, one obtains:
 \begin{eqnarray}
I_2= \int d{\bm r} u_{\alpha\gamma}^2=  \frac{A^2 (1+\beta^2)^3}{16\pi \xi^2\beta^3} \,.
  \label{int2a}
\end{eqnarray}
The sum $\lambda I_1/2+\mu I_2$ gives Eq.\,(\ref{E_mov_vort}) of the main text.

 \section{ }

Apply  $\sum_{\bm G}\int d\omega/2\pi$ to the RHS of   Eq.\,(\ref{lat-mov}):   
 \begin{eqnarray}
   \sum_{\bm a} \delta(\bm r -{\bm v}t-\bm a)=\frac{B}{\phi_0}\sum_{\bm G}
   e^{i{\bm G}( \bm r -\bm v t) }\nonumber\\
  = \frac{2\pi B}{\phi_0}\sum_{\bm G} e^{i{\bm G}( \bm r) }
  \int_{-\infty}^\infty d\omega e^{-i\omega t} \delta(\omega-\bm G \bm v)\,.
   \label{RHS}
\end{eqnarray}
Hence, one has:
 \begin{eqnarray}
&&  \sum_{\bm a} \delta(\bm r -{\bm v}t-\bm a) 
   =\frac{2\pi B}{\phi_0}\sum_{\bm G}
   \int_{-\infty}^\infty \frac{d\omega}{2\pi}  \delta(\omega-\bm G \bm v)e^{i(\bm G \bm r-\omega t)}. \nonumber\\
&&   \label{RHSb}
\end{eqnarray}

Further, one uses 
 \begin{eqnarray}
  \frac{  B}{\phi_0}=  \frac{2\pi  B}{\phi_0}\sum_{\bm G}
   \int_{-\infty}^\infty \frac{d\omega}{2\pi} \delta_{\bm G,0} \delta(\omega)e^{i(\bm G \bm r-\omega t)}.\qquad
   \label{RHSc}
\end{eqnarray}
to obtain Eqs.\,(\ref{chi-v}) and (\ref{chi0}).

\end{document}